\documentclass[a4paper,12pt]{amsproc}

\usepackage{amsaddr}
\usepackage{amsfonts}
\usepackage{amssymb}
\usepackage[colorlinks=true]{hyperref}
\newcommand{\idiv}{\div}
\newcommand{\ld}{\log_2}
\newcommand{\bef}{H_\mathrm{b}}
\newcommand{\prob}[1]{{\mathrm{p}\left[ #1 \right]}}
\newcommand{\integ}{\mathbb{Z}}
\newcommand{\mtxt}[1]{\quad\mbox{\rm #1}\quad}
\newcommand{\mwith}{\mtxt{with}}

\newcommand{\mfor}{\mtxt{for}}
\newcommand{\mand}{\mtxt{and}}

\begin{document}
\title[Saving fractional bits]{Saving fractional bits: A practical entropy efficient code for fair die rolls}
\author{Bernhard \"Omer, Christoph Pacher}
\address{Digital Safety \& Security Department, AIT Austrian Institute of Technology \\
	Donau-City-Stra\ss e 1, A-1020 Vienna, Austria}
	\email{bernhard.oemer@ait.ac.at} \email{christoph.pacher@ait.ac.at}

\begin{abstract}
We give an implementation of an algorithm that uses fair coin flips to simulate fair rolls of an $n$-sided die. A register plays the role of an entropy pool and holds entropy that is generated as a by-product during each die roll and that is usually discarded. The entropy stored in this register is completely reused during the next rolls. Consequently, we can achieve an almost negligible loss of entropy per roll. The algorithm allows to change the number of sides of the die in each round. We prove that the entropy loss is monotone decreasing with increasing entropy pool size (register length).
\end{abstract}

\maketitle

\section{Introduction}

In the digital world, random numbers, be it from pseudo-RNGs, an entropy collection device like Unix' {\tt /dev/random} or a hardware random source, usually come in the form of coin flips i.e. as words of binary digits.
Often, randomness is required to make an unbiased choice between multiple alternatives i.e. to draw a value $r$ from $\integ_n$. 

A standard way to approximate a {\it fair die roll} $D_n$ of {\it range} $n$ with
$$\prob{D_n=r}=\frac{1}{n} \mfor r\in\integ_n $$
is to take a word $X$ of $w$ random bits with $w\gg\ld(n)$ and use $D_n'=X \mod n$ instead of $D_n$. However, this is only exact if $n$ happens to be a power of $2$; otherwise $D_n'$ will show a slight bias towards values $r'<2^w \mod n$.

In the latter case, getting a fair die roll can take more than one try: A straightforward implementation might reroll $r\gets X$ until $r<n$. If the word size of $X$ is tightly fitted, i.e. $w=\lceil \ld(n) \rceil$ this procedure will on average require less than two tries and consume less than $2+2\ld(n)$ bit of entropy. Still, drawing from a short Poker deck with Joker ($D_{33}$) would use on average $\left( \ld (64) \right) / \frac{33}{64}=11.636$ bit to produce $\ld(33)=5.044$ bit of output for an entropy efficiency of only $\eta = 43.35\%$.

This is not a problem if entropy is both cheap and unbounded but can make a difference if your random bits are slowly harvested from I/O operations, derived from complex cryptographic protocols or brought to you by armed guys in dark suits.

\subsection{Related Work}

Lumbroso \cite{Lumbroso2013} has given a run-time efficient algorithm to draw random discrete uniform variables within a given range of size $n$ (corresponding to the outcome of a fair roll of an $n$-sided die) from a source assumed to produce independent and unbiased random bits (unbiased coin flips). This recent work contains also a summary of previous related work so that we refrain from giving another overview. However, we want to mention that also the problem of generating fair die rolls from \emph{biased} random bits (biased coin-flips) has been studied \cite{Gargano1999}.

\subsection{Our contribution}

We give another implementation of an algorithm that uses fair coin flips to simulate fair rolls of an $n$-sided die. A register holds entropy that is generated as a by-product during a die roll and that is usually discarded. One interesting property of this algorithm is that we can completely reuse the entropy stored in this register during the next rolls; one register is enough for the accumulation of entropy for an arbitrary number of rolls. Consequently, we can achieve an almost negligible loss of entropy per roll. Another interesting property is that the algorithm allows to change the number of sides of the die in each round without further entropy loss. Furthermore we perform an analysis and show that increasing the entropy pool size (the register length) reduces the entropy loss of the algorithm.

\section{Plugging the entropy leaks}

In the above algorithm, entropy gets wasted in two places: Firstly you have the {\it offcut} which derives from your entropy being only available in bit-sized packets and secondly, you have the {\it discard} when your drawn binary word does not fall into the required range.
Both effects can be mitigated but none can be totally avoided.

Let us deal with the offcut first. If the ranges $n_i$ of the rolls are known in advance, then you can treat them as a single roll $D_n$ over the product range $n=\prod n_i$ and interpret the outcome as a mixed radix number. This limits the total offcut to less then one bit and is optimal.

But what if the ranges are not known in advance or in fact depend on the outcome of previous rolls? In that case, we can simply make one up and store the result for later use. For this to be of any use, we need a method to store entropy in non-integral multiples of bits. This sounds more esoteric as it is --- after all, we know that an ordinary cubic die under a cup holds exactly $\ld(6)=2.585$ bit of information. The state of a  die is thus easily stored by the pair $(m,t)$ with $m$ being the number of faces and $t$ being the actual zero-based face value.

So instead of a guess-tape with $l$ binary digits, we treat our initial entropy pool as a $2^l$-faced die with the state $s_0=(m_0,t_0)$ with $m_0=2^l$ and $t_0$ being the tape-content interpreted as binary number. If a $D_n$ is to be drawn from the pool in state $s=(m,t)$, then $s$ is updated by a complimentary draw of $D_k'$ with $k=m\idiv n=\lfloor m/n \rfloor$, so
\begin{equation}\label{sred1}
  s'=(m \idiv n,t \idiv n) \mand D_n \to t \mod n \mfor t<nk.
\end{equation}

This method is exactly equivalent to the case when the ranges are known in advance and will yield the same results with the same chance of success when run on identical entropy pools. Practically, one might limit the initial size of the pool (e.g. to one wordlength of the CPU) and ``refill'' after each draw which will incur only very minor penalities in terms of offcut if $m\gg n$. A new bit $b$ can simply be shifted into the state i.e $(m,t)\to(2m,2t+b)$.

Note that in (\ref{sred1}) we have not defined $s'$ when $t\ge nk$. It is clear that we cannot produce any output in that case (doing so would bias both $D_n$ and $D_k'$), so the die roll failed and has to be redrawn. But what about the state? Has it to be discarded as well? 

Not completly. While we failed to produce an unbiased $D_n \times D_k'$ roll, not all is lost: By checking for overflow we learned not only that $t<nk$ but also that $t\ge nk$; moreover we know that $t<m$. We have not learned about or acted upon any other information from the entropy pool, thus $t$ is still uniformly distributed in the integer range $[nk,m)$ which is perfectly good entropy left to recycle. All we need to do is transform the range, so
\begin{equation}\label{sred2}
  s'=(m-nk,t-nk) \mand D_n \to \mathbf{undef} \mfor t\ge nk.
\end{equation}

\section{Limits of Recycling}

Since we can recycle some of the discard, we might want to reconsider our original decision to set $k=m\idiv n$ and thus to the maximum possible value. Maybe a smaller value of $k$ or eliminating the complimentary draw $D_k'$ completely (i.e. setting $k=1$) would work even better? Also, should the entropy pool be as huge as feasible or be kept small, by only refilling to accomodate the current die roll.

To answer this question, let us calculate the exact amount of waste $w$ per iteration by comparing the (Shannon-) entropies. Before, we have $ S=\ld m $ and afterwards either $S_1'=\ld k$ plus the entropy of the output $S_o=\ld n$ or $S_2'=\ld(m-nk)$ in case of a discard. 
The net balance is thus
\begin{equation}\label{entr1}
  w=S-\frac{nk}{m}(S_1'+S_o)-\frac{m-nk}{m}S_2',
\end{equation}
which can be transformed into
\begin{equation}\label{entr2}
  w=\bef(p)=-p \ld p-(1-p) \ld (1-p) \mwith p=\frac{nk}{m}.
\end{equation}

$\bef(p)$ is the binary entropy function, giving the amount of information we learn from flipping a biased coin. The ``coin'' in question was of course our overflow test, so the missing entropy is exactly the amount of information we have learned by checking $t<nk$.

To get the waste per roll $W$, we have to multiply by the expected number of iterations $1/p$ and get
\begin{equation}\label{waste}
  W=\frac{\bef(p)}{p} \mand \frac{\partial W}{\partial p}=\frac{\ld(1-p)}{p^2}<0.
\end{equation}

So $W$ is strictly decreasing with $p$ and thus with $k$, so setting $k=m\idiv n$ is indeed optimal and we learn that minimizing the waste in the first place beats recycling.\footnote{If a wasteful implementation is done anyway, then the entropy pool should always be kept as small as possible so that $p=\frac{n}{m}>\frac{1}{2}$.}
For statistically independent $m,n$ with $m\gg n$, we can assume $p\approx 1-\frac{n}{2m}$ so it makes sense to keep the entropy pool big. For this case, we can estimate the efficiency 
\begin{equation}\label{efficiency}
  \eta=\frac{S_o}{S_o+W}\approx 1-\frac{n}{2m}\frac{1+\ln 2+\ln m-\ln n}{\ln n}.
\end{equation}

\newpage
\section{Sample Implementation}

The C-code below implements an efficient die using an external entropy souce \verb=dev_random=. 
It uses two integer divisions per iteration, using both quotient and remainder.

\begin{center}
\small
\begin{verbatim}
unsigned die(unsigned n) {
  static unsigned long m = 1, t = 0;   /* init entropy reservoir */
  unsigned long k, l, nk, r;
  
  for(;;) {
    while(m <= ULONG_MAX >> 8) {       /* top off reservoir      */
      m <<= 8;
      t = (t << 8) | getc(dev_random); /* read entropy byte      */
    }
    k = m / n;            /* max out range of complimentary draw */
    l = m % n;  
    nk = m - l;
    if(t < nk) {          /* do we have a valid n x k draw?      */
      r = t % n; 
      m = k;  t /= n;     /* yes: reduce state and return result */
      return r;
    }
    m = l;  t -= nk;      /* no: recycle the discard and repeat  */
  }
}
\end{verbatim}
\end{center}

\end{document}